\documentclass[sigconf,natbib=true]{acmart}
\usepackage{amsmath,amsfonts}
\usepackage{graphicx}
\usepackage{textcomp}
\usepackage{xcolor}
\usepackage{booktabs}
\usepackage{multirow}
\usepackage{xspace}
\usepackage{balance}
\usepackage{bm}
\usepackage{url}
\usepackage{hyperref}
\usepackage{marvosym}
\usepackage{multicol}
\usepackage{xspace}
\usepackage[flushleft]{threeparttable}
\usepackage{adjustbox}
\usepackage{graphicx}
\usepackage{subcaption}
\usepackage{enumitem}
\usepackage[ruled,linesnumbered]{algorithm2e}

\definecolor{c1}{rgb}{1.000, 0.753, 0.796}
\definecolor{c2}{rgb}{0.980, 0.502, 0.447}
\definecolor{c3}{rgb}{1.000, 0.498, 0.314}
\definecolor{c4}{rgb}{0.863, 0.078, 0.235}
\definecolor{c5}{rgb}{0.863, 0.078, 0.235}
\definecolor{c6}{rgb}{0.545, 0.000, 0.000}
\definecolor{lightgray}{rgb}{0.9, 0.9, 0.9}
\definecolor{c1}{rgb}{0.9, 0.9, 0.9}

\AtBeginDocument{%
  \providecommand\BibTeX{{%
    \normalfont B\kern-0.5em{\scshape i\kern-0.25em b}\kern-0.8em\TeX}}}

\setcopyright{acmlicensed}
\copyrightyear{2018}
\acmYear{2018}
\acmDOI{XXXXXXX.XXXXXXX}

\acmConference[Conference acronym 'XX]{Make sure to enter the correct
  conference title from your rights confirmation emai}{June 03--05,
  2018}{Woodstock, NY}
\acmBooktitle{Woodstock '18: ACM Symposium on Neural Gaze Detection,
 June 03--05, 2018, Woodstock, NY} 
\acmISBN{978-1-4503-XXXX-X/18/06}

\begin{document}


\author{Xinzhou Jin$^{1\dagger}$, Jintang Li$^{1}$, Liang Chen$^{1}$\textsuperscript{\Letter}, Chenyun Yu$^{1}$, Yuanzhen Xie$^{2}$\\ Tao Xie$^{2}$, Chengxiang Zhuo$^{2}$, Zang Li$^{2}$, Zibin Zheng$^{1}$}

\thanks{$\dagger$ Work done during an intern at Tencent.}
\thanks{\textsuperscript{\Letter} Corresponding author.}

\affiliation{
  \institution{$^1$Sun Yat-sen University, China}
  \institution{$^2$Platform and Content Group, Tencent, China}
  \country{}
}

\affiliation{
  \institution{jinxzh5@mail2.sysu.edu.cn, chenliang6@mail.sysu.edu.cn}
  \country{ }
}

\newcommand{\ie}{\emph{i.e.,}\xspace}
\newcommand{\eg}{\emph{e.g.,}\xspace}
\newcommand{\aka}{\emph{a.k.a.,}\xspace}
\newcommand{\etal}{\emph{et al.}\xspace}
\newcommand{\paratitle}[1]{\vspace{1.5ex}\noindent\textbf{#1}}
\newcommand{\wrt}{w.r.t.\xspace}
\newcommand{\ignore}[1]{}

\newcommand{\tba}{\textcolor{red}{xxx }}
\newcommand{\ours}{L$^2$CL }
\newcommand{\nours}{L$^2$CL}
\newcommand{\plot}{l2cl}

\renewcommand{\authors}{Xinzhou Jin, Jintang Li, Liang Chen, Chenyun Yu, Yuanzhen Xie, Tao Xie, Chengxiang Zhuo, Zang Li, Zibin Zheng}
\renewcommand{\shortauthors}{Jin et al.}

\begin{abstract}
Graph neural networks (GNNs) have recently emerged as an effective approach to model neighborhood signals in collaborative filtering. Towards this research line, graph contrastive learning (GCL) demonstrates robust capabilities to address the supervision label shortage issue through generating massive self-supervised signals.
Despite its effectiveness, GCL for recommendation suffers seriously from two main challenges: i) GCL relies on graph augmentation to generate semantically different views for contrasting, which could potentially disrupt key information and introduce unwanted noise;  ii) current works for GCL primarily focus on contrasting representations using sophisticated networks architecture (usually deep) to capture high-order interactions, which leads to increased computational complexity
and suboptimal training efficiency.
To this end, we propose \nours, a principled \underline{\textbf{L}}ayer-to-\underline{\textbf{L}}ayer \underline{\textbf{C}}ontrastive \underline{\textbf{L}}earning framework that contrasts representations from different layers.
By aligning the semantic similarities between different layers, \ours enables the learning of complex structural relationships and gets rid of the noise perturbation in stochastic data augmentation. 
Surprisingly, we find that \nours, using only \textit{one-hop} contrastive learning paradigm, is able to capture intrinsic semantic structures and improve the quality of node representation, leading to a simple yet effective architecture. 
We also provide theoretical guarantees for \ours in minimizing task-irrelevant information.
Extensive experiments on five real-world datasets demonstrate the superiority of our model over various state-of-the-art collaborative filtering methods.
Our code is available at \url{https://github.com/downeykking/L2CL}.
\end{abstract}





\title{\nours: Embarrassingly Simple Layer-to-Layer Contrastive Learning for Graph Collaborative Filtering}
\maketitle

\vspace{-0.2cm}
\section{Introduction}\label{sec:intro}
Recommender systems are widely used to cope with the problem of information overload, and play a vital role in many online information systems, such as e-commerce platforms ~\cite{smith2017two}, video websites~\cite{covington2016deep}, social media~\cite{godin2013using}. Collaborative filtering (CF) is the core algorithm behind recommender systems, which is based on the assumption that similar users tend to have similar preferences~\cite{rendle2009bpr}. It captures user preferences through implicit feedback records (such as click and view). Traditional CF methods map users' and items' IDs into latent embeddings and utilize matrix factorization~\cite{koren2009matrix, mnih2007probabilistic} to optimize the representation of users and items. 

With the recent advancements in graph representation learning~\cite{kipfsemi, wu2019simplifying}, several works~\cite{vdberg2017graph, ying2018graph, sun2019multi, wang2019neural, he2020lightgcn} introduce graph neural networks (GNNs) to extract local collaborative signals and achieve significant improvements. The GNN-based collaborative filtering method captures rich multi-hop neighborhood information through message passing in the user-item interaction graph.
Despite the popularity and effectiveness of applying GNNs to CF, the learning of high-quality user and item representations is hindered by various challenges, including data sparsity, noisy interactions, and the heavy reliance on sufficient supervision signals.

To address the aforementioned challenges, researchers have shifted their attention to contrastive learning (CL), a promising learning paradigm that enables training on vast amounts of unlabeled data in a self-supervised manner
~\cite{oord2018representation, velivckovic2018deep, sun2019infograph, gao2021simcse}. The primary concept underlying contrastive learning in recommendation is to maximize the alignment between the generated embedding views by contrasting positively defined pairs with their corresponding negative instances.
Therefore, how to design contrastive views to promote representation learning has become the core of CL design.

Early works~\cite{wu2021self, yao2021} apply random augmentation to the interaction graph or perform feature masking on node representations to build contrastive views. Following~\cite{gao2021simcse}, subsequent efforts~\cite{yu2022graph, yu2023xsimgcl} create contrastive pairs through perturbing the learned node representations. Additionally, some methods~\cite{hassani2020contrastive, jiang2023adaptive} generate self-supervised signals through heuristic-guided view construction or fine-grained hypergraph~\cite{hccf2022} definition. 

While the above-mentioned GCL-based CF models have advanced state-of-the-art performance for recommendation, there are two key issues to be addressed: (1) Modifying the original graph structure or node features may potentially harm critical structural information and introduce unexpected noise. The success of heuristic-guided representation-contrast schemes heavily relies on the additional augmentation for generating semantic views (e.g., user clustering, embedding perturbing, hypergraph generating), but in various recommendation scenarios, achieving accurate generation of contrastive views is very challenging, and manual augmentation may unavoidably introduce noisy and irrelevant information for self-supervised learning.
(2) Existing works for GCL in recommendation primarily focus on contrasting representations using sophisticated deep networks architecture to model high-order interactions, 
which overlooks the noisy information may be amplified by the stacking of multiple convolutional layers in deep GCL architectures, which raises the risk of incorporating misleading self-supervised signals and results in a higher computation burden.

In light of the above challenges, we revisit the GCL paradigm for recommendation and propose a novel \textbf{layer-level} (layer-to-layer in detail) graph contrastive learning framework, \ours, in place of the predominant \textbf{view-level} paradigm.
In particular, following~\cite{lin2022ncl}, given an $L$-layer GNN, we employ the output from the $l$-th layer as the representations of $l$-hop neighbors for a node and devise a structure-aware contrastive learning objective, 
which aims at aligning the semantic similarities between a node and its structural neighbors in different layers. 
The layer-to-layer contrasts eliminate the necessity for manually crafted view generate techniques and mitigate the potential noise introduced by random augmentation.

By treating all meaningful structural neighbors as positive instances, 
we systematically unify \ours with various contrastive learning schemes through different layer contrasts. In this framework, the learned node representations from different layers include different interactive collaborative signals, thus employing these strengthened relationships for constructing contrastive pairs possesses robust structure-awareness capabilities. 

Furthermore, motivated by the theory of relevant information maximization and irrelevant information minimization, we perform \ours in only utilizing a 1-layer GNN and 1-hop structural neighbors. Instead of utilizing high-order information, intrinsic characteristics of the representations towards downstream tasks can be well preserved in our 1-hop \ours. This enables robustness against 
noisy information which may be propagated and stored in deep graph convolution.

The main contributions of this paper are summarized as follows:
\begin{itemize}[leftmargin=*]
\item We propose and unify a layer-to-layer contrastive learning paradigm, which heavily exploits the graph structure information and is capable of mining implicit collaboration signals of neighbors from different hierarchical layers without requiring graph-level augmentation and hand-crafted view generation.
\item We further simplify \ours by only considering one-hop structural neighbors as enhanced positive pair neighbors. Utilizing a single-layer GNN enhances the model's efficiency and mitigates the suboptimal performance caused by excessive smoothing of node representations.
\item We provide a theoretical explanation for the effectiveness of \nours, which may offer new insights for contrastive learning in recommendation.
\item Extensive experimental results on five real-world datasets demonstrate the superiority of our proposed method over the existing CF methods in both effectiveness and efficiency.
\end{itemize}
\section{preliminaries}\label{sec:preliminaries}

\subsection{Notation and Problem Definition}

In this section, we present some useful notations and definitions. Given the user set $\mathcal{U} (|\mathcal{U}|=M)$ and item set $\mathcal{I} (|\mathcal{I}|=N)$, we focus on the implicit feedback problem within the context of recommender systems. The implicit feedback matrix is denoted as $R \in\{0,1\}^{M \times N}$. Specifically, each entry $R_{u, i}$ indicates whether the user $u$ is connected to item $i$, with a value of 1 representing a connection and 0 otherwise.

Graph collaborative filtering methods typically organize the interaction data $R$ into the pattern of an user-item bipartite graph $\mathcal{G}=\{\mathcal{V}, \mathcal{E}\}$, where $\mathcal{V}=\{\mathcal{U} \cup \mathcal{I}\}$ denotes the set of nodes, and $\mathcal{E}=\left\{(u, i) \mid u \in \mathcal{U}, i \in \mathcal{I}, R_{u, i}=1\right\}$ denotes the set of edges. The adjacent matrix $\mathrm{A}$ is  defined as follows:
\begin{equation}
\mathrm{A}=\left[\begin{array}{cc}
0^{M \times M} & \mathbf{R} \\
\mathbf{R}^T & 0^{N \times N}
\end{array}\right].
\end{equation}
Given the initialized node embeddings ${E}^{(0)}$, 
GNNs update node representation through aggregating messages of its
neighbors. As such, the core of the graph-based CF paradigm consists of two steps: (1) message propagation; and (2) node representation aggregation.
Thus, the layer of GNNs can be defined as:
\begin{equation}
E_u^{(l)}=\operatorname{COMBINE}(\{E_u^{(l-1)}, \operatorname{AGGREGATE}^{(l)}(\{E_i^{(l-1)}: i \in \mathcal{N}_u\}\}), 
\end{equation}

\begin{equation}
E_u={\text {READOUT}}([E_u^{(0)}, E_u^{(1)}, \ldots, E_u^{(L)}]),
\end{equation}
where $\mathcal{N}_u$ denotes the neighborhood set of node $u$, and $L$ denotes the number layers of graph convolution.

Subsequently, we derive scores for all unobserved user-item pairs using the inner product of the final representations of the user and item, denoted as $y_{ui} = e_{u}^{T}e_i$, where $e_{u}$ and $e_i$ represent the representations of user $u$ and item $i$, respectively. The item with the highest score is recommended to the user. 

To optimize, we use the Bayesian personalized ranking (BPR)
loss \cite{rendle2009bpr} as follow:
\begin{equation}
\label{eqn:4}
\mathcal{L}_{rec}=-\sum_{u=1}^{|\mathcal{U}|} \sum_{i \in \mathcal{N}_u} \sum_{j \notin \mathcal{N}_u} \log \sigma\left({e}_u^T e_i-{e}_u^T {e}_j\right),
\end{equation}  
where $j$ represents a randomly selected negative item that the user has not previously interacted with.

\subsection{GCL for Recommendation}
The core idea of contrastive learning~\cite{chen2020simple} is to 
pull different variants of the same instance (e.g., user, item) closer together in the embedding space while pushing variants of different instances further apart. 
In recommendation, these instances are generated based on the various augmentation methods to offer semantically different views for contrasting, which can be seen as \textbf{view-level} GCL.

In GCL, the efficient graph encoder, LightGCN~\cite{he2020lightgcn} is widely used as the backbone. It achieves superior performance by removing redundant operations, including transformation matrices and activation functions. Its formulas for message passing and embedding propagation with initialized $e_u^{(0)}$, $e_i^{(0)}$ are as follows:

\begin{equation}
\label{lightgcn1}
    {{e_{u}}}^{(l)}=\sum_{i \in \mathcal{N}_u} \frac{1}{\sqrt{\left|\mathcal{N}_u\right|\left|\mathcal{N}_i\right|}} {{e_{i}}}^{(l-1)},
\end{equation}
\begin{equation}
\label{lightgcn2}
    {{e_{i}}}^{(l)}=\sum_{e \in \mathcal{N}_i} \frac{1}{\sqrt{\left|\mathcal{N}_i\right|\left|\mathcal{N}_u\right|}} {{e_{u}}}^{(l-1)}.
\end{equation}

Specifically, each data point $e_u^{(0)}$ is augmented using a random transformation to obtain $e_u^{(0)'}$. These original and augmented data points are processed through LightGCN to generate node representations at each layer and subsequently, a weighted sum is performed to obtain the final output:
\begin{equation}
    e_u=\frac{1}{L+1} \sum_{l=0}^L {e_u}^{(l)}, \quad e_i=\frac{1}{L+1} \sum_{l=0}^L {e_i}^{(l)}.
\end{equation}

By maximizing the consistency among positive samples and minimizing the consistency among negative samples, the contrastive loss based on InfoNCE~\cite{oord2018representation} can be expressed as:
\begin{equation}
    \mathcal L_{cl}^{U} = \sum_{u\in \mathcal{U}}-\log \frac{\exp(s(\mathbf {e}_u, \mathbf {e}_u^{'})/\tau)}{\sum_{k\in \mathcal{U}} \exp(s(\mathbf {e}_u, \mathbf {e}_k^{'})/\tau)} ,
\end{equation}
where $s(\cdot)$ represents cosine similarity and $\tau$ is the temperature. Similarly, item-side contrastive learning is performed analogously.

\section{METHODOLOGY}\label{sec:method}

\subsection{\nours: Unifying Layer-to-Layer Contrast in Graph Collaborative Filtering}
The view-level contrastive learning relies on generating accurate and semantic contrastive views, but it is challenging for such view generation because additional augmentation may introduce irrelevant information and noise. Revisiting the message-passing mechanism in GNNs, the origin collaborative signals (e.g. observed interactions) are propagated to different layers through extended high-order paths, which inspires us that different hop nodes may be potential structural neighbors and the outputs of the different layers may share the similarity in embedding space.
To this end, we propose the layer-to-layer contrastive learning paradigm that contrasts representations from different layers.

We utilize LightGCN as graph convolution backbone and obtain node embeddings $e_u^{(0)}, e_u^{(1)}, \ldots, e_u^{(L)}$ at each layer like Eq. (\ref{lightgcn1}, \ref{lightgcn2}).

Considering the output of different layers as node representations and using them to construct positive contrasting pairs, the layer-to-layer contrasts can be further categorized into homogeneous and heterogeneous. Homogeneous indicates contrasting nodes from the same type (e.g., user-user) with different layers. The generic contrastive learning loss for the user side is defined as:
\begin{equation}
    \mathcal L_{cl}^{U} = \sum_{u\in \mathcal{U}}-\log \frac{\exp(s(\mathbf {e}_u^{(m)}, \mathbf {e}_u^{(n)})/\tau)}{\sum_{k\in \mathcal{U}} \exp(s(\mathbf {e}_u^{(m)}, \mathbf {e}_k^{(n)})/\tau)} ,
\end{equation}
where $0 \leq m,n  <=L	$, the same for the item side.

Besides, for users, their updated representation is aggregated through directly connected items. Therefore, despite users and items belonging to different classes of heterogeneous node types, their representation spaces may still be similar. so the contrastive learning loss for heterogeneous nodes is defined as:
\begin{equation} \label{cl_loss}
    \mathcal L_{cl}^{U} = \sum_{i \in \mathcal{I}, \ R_{u, i}=1}-\log \frac{\exp(s(\mathbf {e}_i^{(m)}, \mathbf {e}_u^{(n)})/\tau)}{\sum_{k\in \mathcal{U}} \exp(s(\mathbf {e}_i^{(m)}, \mathbf {e}_{k}^{(n)})/\tau)}, 
\end{equation}

\begin{figure}[!h]
	\centering
	\includegraphics[width=0.45\textwidth]{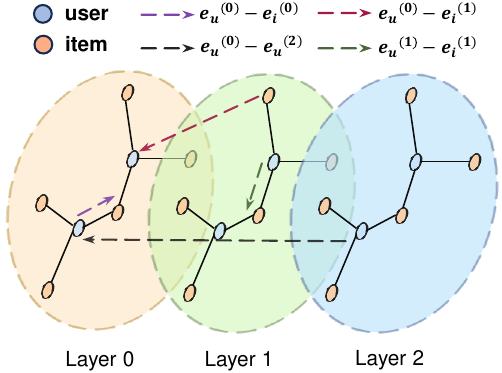}
	\caption{Different layer-to-layer contrast.}
	\label{fig:unify}
\end{figure}

To reduce the excessive number of contrastive pair combinations, we step forward and unify five instances in layer-to-layer contrast. We denote $e_u^{(0)}-e_i^{(0)}$ as performing contrastive learning in heterogeneous nodes while the embedding of user $u$ at layer 0 is anchor, and the embedding of directly connected item $i$ at layer 0 is a positive instance. Given a two-layer LightGCN, Figure~\ref{fig:unify} illustrates the detail for different layer contrast.

The contrasts $e_u^{(0)}-e_i^{(0)}$ and $e_u^{(1)}-e_i^{(1)}$ are intuitive since the contrasting pairs have explicit interaction information in the original bipartite graph. 
$e_u^{(2)}$ is obtained through embedding propagation of $e_u^{(0)}$, which can be seen as the aggregation of potential structural neighbors similar to~\cite{lin2022ncl}.
The contrast $e_u^{(0)}-e_i^{(1)}$ utilizes one-hop neighbor information from explicit interactions, for $e_i^{(1)}$ is the direct aggregation of $e_u^{(0)}$ and other user nodes. The contrast $e_u^{(0)}-\sum_{u=1}^3{e_u^{(i)}}$ implicitly captures collaborative signals from different layers through weighted aggregation.
In summary, \ours brings new insight into generating meaningful positive samples. However, under the unifying framework, seeking effective layer contrast with sufficient collaborative signals is supposed to be explored further, for different layer-to-layer contrasts will bring different collaborative signals. 
Therefore, a natural question arises: 
\begin{center}
\textit{What is the best layer-to-layer paradigm for recommendation?}
\end{center}

\subsection{One-Hop Contrastive Learning}\label{sec: one-hop}
To answer the question, 
We conduct the above layer-to-layer contrast experiments on two public benchmarks, Yelp and Books~\cite{he2020lightgcn}. the results\footnote{For the sake of brevity, we omit the item side here for presentation purposes. We also contrast $e_i^{(0)}-e_u^{(1)}$ while performing the contrast of $e_u^{(0)}-e_i^{(1)}$. More experimental details see Section ~\ref{sec:sec4-exp}.} is shown in Table~\ref{tab:exp-unify}.
\begin{table}[!ht]
  \centering
  \small
  \caption{Performance comparison of different \ours variants.}
  \label{tab:exp-unify}
  \begin{tabular}{ccccc}
    \toprule
    \multirow{2}{*}{Method} & \multicolumn{2}{c}{Yelp} & \multicolumn{2}{c}{Books} \\
    \cmidrule(lr){2-4}\cmidrule(lr){4-5}
    & Recall@10 & NDCG@10 & Recall@10 & NDCG@10 \\
    \midrule
    LightGCN        & 0.0730     & 0.0520     & 0.0797     & 0.0565     \\
    $e_u^{(0)}-e_i^{(0)}$      & 0.0975     & 0.0761     & 0.0928     & 0.0671     \\
    $e_u^{(1)}-e_i^{(1)}$      & 0.0932     & 0.0702     & 0.1074     & 0.0778     \\
    $e_u^{(0)}-e_i^{(1)}$                & 0.0995     & 0.0745     & 0.1067     & 0.0767     \\
    $e_u^{(0)}-e_u^{(2)}$     & 0.0814     & 0.0606     & 0.0819     & 0.0586     \\
    $e_u^{(0)}-\sum_{i=1}^3{e_u^{(i)}}$   & 0.0809     & 0.0615     & 0.0831     & 0.0598     \\
    \bottomrule
  \end{tabular}
\end{table}

In Table~\ref{tab:exp-unify}, the gains from contrasting using 1-hop structural information, such as $e_u^{(0)}-e_i^{(0)}$ and $e_u^{(0)}-e_i^{(1)}$, far outweigh the benefits from implicit high-order information like $e_u^{(0)}-e_u^{(2)}$. This insight inspires us that in GCL-based models, the information derived from direct observation interactions has not been fully exploited and still holds significant utility.

The fundamental assumption of collaborative filtering is that similar users tend to have similar preferences. In detail, When user $u$ directly connects item $i$, it may share similar interests with those users who have historically interacted with the same item $i$. Let $e_u^{(0)}$, $e_i^{(0)}$ denote the initialized user $u$ and its connected item $i$ respectively, after one-layer graph convolution, $e_i^{(1)}$  aggregates users' embeddings (e.g., $e_u^{(0)}$) who have consumed it, which can be seen as the fusion information of user $u$. This indicates that $e_u^{(0)}$ and $e_i^{(1)}$ should be similar in the embedding space, which aligns intending to contrast $e_u^{(0)}$ and $e_i^{(1)}$.

Therefore, we utilize $e_u^{(0)}$ and $e_i^{(1)}$, as well as $e_i^{(0)}$ and $e_u^{(1)}$, to construct our contrastive learning loss function:

\begin{equation} \label{cl_loss}
    \mathcal L_{cl}^{U} = \sum_{i \in \mathcal{I}, \ R_{u, i}=1}-\log \frac{\exp(s(\mathbf {e}_i^{(1)}, \mathbf {e}_u^{(0)})/\tau)}{\sum_{k\in \mathcal{U}} \exp(s(\mathbf {e}_i^{(1)}, \mathbf {e}_{k}^{(0)})/\tau)}, 
\end{equation}
where ${e}_i^{(1)}$ is output of GNN at layer 1. In one batch, item $i$ is the direct \textbf{one-hop neighbors} of user $u$ in a bipartite graph, whose representation can be obtained like:
\begin{equation}
{{e_{i}}}^{(1)}=\sum_{u \in \mathcal{N}_i} \frac{1}{\sqrt{\left|\mathcal{N}_u\right|\left|\mathcal{N}_i\right|}} {{e_{u}}}^{(0)}.
\end{equation}

In a similar way, the one-hop structural contrastive learning loss of the item side 
can be obtained as:
\begin{equation} \label{cl_loss}
    \mathcal L_{cl}^{I} = \sum_{u \in \mathcal{U}, \ R_{u, i}=1}-\log \frac{\exp(s(\mathbf {e}_u^{(1)}, \mathbf {e}_i^{(0)})/\tau)}{\sum_{k\in \mathcal{I}} \exp(s(\mathbf {e}_u^{(1)}, \mathbf {e}_{k}^{(0)})/\tau)}.
\end{equation}

Overall, the final contrastive loss is the weighted sum of the user-side loss and item-side
loss with the coefficient $\alpha$ to balance two contrastive losses:
\begin{equation} \label{cl_loss_whole}
    \mathcal L_{cl} = \alpha\mathcal L_{cl}^{U} + (1-\alpha)\mathcal L_{cl}^{I}.
\end{equation}

\subsection{Theoretical Guarantees}
\begin{figure}[!h]
	\centering
	\includegraphics[width=0.35\textwidth]{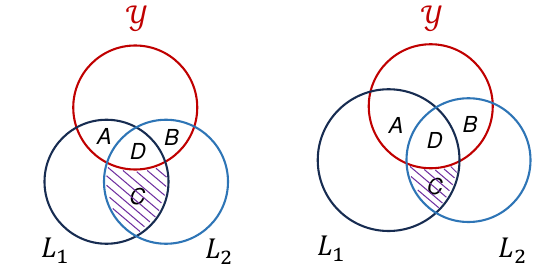}
	\caption{The left is an illustration of information in self-supervised learning.
 $L_1$, $L_2$, and $\mathcal{Y}$ represent the information of two layers and downstream tasks, respectively. The right is the asymmetric layer contrast strategy, which can alleviate the task-irrelevant noises in GCL (area C).}
	\label{fig:information}
\end{figure}

To illustrate the effectiveness of one-hop layer-to-layer contrast, we shift our focus back to the self-supervised learning objective driven by information theories. As shown in Figure~\ref{fig:information}, where the scale of each area reflects the amount of information, the learned representations through CL include both task-relevant information (area D) and task-irrelevant noises (area C). 
The information-theoretic characterizations~\cite{tian2020makes, Wang_2022_CVPR} indicate that a well-adapted self-supervised learning method for the downstream task should \textit{extract task-relevant information} in the latent representation and \textit{discard task-irrelevant information},  
so the purpose of \ours is to expand information in area D and limit information in area C.
\subsubsection*{\textbf{Maximizing task-relevant information}}

Considering one-layer LightGCN, the calculation of the similarity $s_{ij}$ between any user $u_i$ and item $i_j$ is expanded~\cite{anonymous2023how} by introducing three additional similarity terms as inductive biases:
the similarity between users who purchase the same item (e.g., ($\mathbf{u}_i^{\top} \cdot \mathbf{u}_k$)), the similarity between items purchased by the same user (e.g., $(\mathbf{i}_k^{\top} \cdot \mathbf{i}_j)$), and the similarity between neighbors observed in interactions (e.g., $(\mathbf{i}_k^{\top} \cdot \mathbf{u}_k)$). In the collaborative filtering scenario, the downstream task aims to recommend items to users they are most likely to interact with, and the BPR is used to update the similarity between directly connected user-item pairs. Considering the results in Table~\ref{tab:exp-unify}, building contrastive loss for direct 1-hop user-item nodes leads to greater improvements, suggesting that the value of first-order interaction information in the recommendation task has not been fully exploited. Therefore, modeling these three inductive biases into the contrastive learning can better align with downstream tasks and result in expanded area D. The contrasting pairs modeling these three inductive biases correspond to $e_u^{(0)}-e_i^{(1)}$, $e_i^{(0)}-e_u^{(1)}$, and $e_u^{(0)}-e_i^{(0)}$. The first two are the objects used in one-hop  \nours.
\subsubsection*{\textbf{Minimizing task-irrelevant information}}
The layer-wise message passing in GNN introduces a lot of redundancy, and these redundancies are preserved in the deep latent representations of adjacent nodes in the form of task-irrelevant information. After $k$-layer message passing of GNN, the output latent representation contains aggregated information from $k$-hop subgraphs. For two adjacent nodes, their $k$-hop subgraphs have a significant overlap, resulting in a substantial correlation between the representations.

Let $N^k_{uv}$ denote the overlap size of the subgraphs $X$ and $Y$, $N_k$ represent the maximum size of all overlapping subgraphs. $T$ denote downstream task. Suppose the features of each node in the graph are independently and identically sampled from a Gaussian distribution $G(0, \gamma)$, where $\gamma$ lies in the range [-1, 1]. Proposition 1 in ~\cite{li2023maskgae} provides a lower bound for task-irrelevant information:
\begin{equation}
I(X;Y|T) \geq \frac{\left(\mathbb{E}\left[N_{u v}^k\right]\right)^2}{N_k} \gamma^2.
\end{equation}
This lower bound indicates that the task-irrelevant information scales almost
linearly with the size of overlapping subgraphs. Compared to deep graph convolution, shallow GNN always has a lower subgraph overlapping size~\cite{li2023maskgae}, which is consistent with our target to only utilize 1-hop structural neighbors to keep lower task-irrelevant information. 
Besides, Figure~\ref{fig:information} indicates that employing an asymmetric layer contrast strategy is better for decreasing the scale of area C, which alleviates the task-irrelevant noises in GCL. Therefore, $e_u^{(0)}-e_i^{(1)}$ can reduce more redundancy in the self-supervised signal compared with $e_u^{(0)}-e_i^{(0)}$ and $e_u^{(1)}-e_i^{(1)}$.

\subsection{Model Training}
\begin{figure}[ht]
	\centering
	\includegraphics[width=0.48\textwidth]{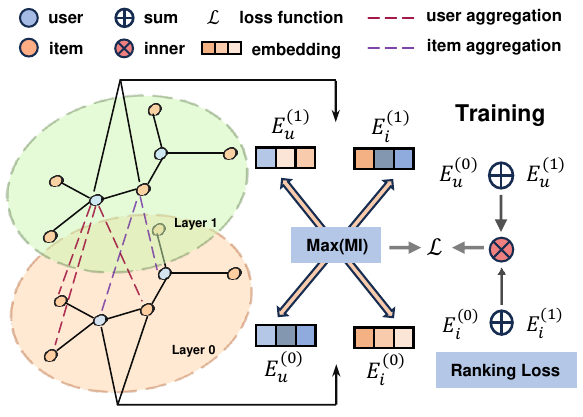}
	\caption{The illustration of our proposed \nours, which jointly optimize one-hop contrastive learning loss and BPR loss.}
	\label{fig:model}
\end{figure}

To learn more meaningful node representations for \nours, we use a multi-task training strategy to jointly optimize the traditional recommendation task (Eq. (\ref{eqn:4})) and the contrastive learning task\footnote{Unless stated otherwise, the contrastive objective for \ours in the following section is the one-hop neighbors in Section~\ref{sec: one-hop}.} (Eq. (\ref{cl_loss_whole})):
\begin{equation}
\label{eqn:12}
    \mathcal{L} = \mathcal{L}_{rec} + \lambda_1\mathcal{L}_{cl} + \lambda_2||{\mathbf{E}}^{(0)}||^2,
\end{equation}
where $\lambda_1$ and $\lambda_2$ are hyperparameters that control the strength of contrastive learning and $L_2$ regularization respectively. 
The whole workflow of one-hop \ours is shown in Figure~\ref{fig:model}.

\section{Experiments} \label{sec:sec4-exp}

To evaluate the effectiveness of the proposed \ours in recommendation tasks, we perform comprehensive experiments and present detailed results and analysis. We aim to answer the following research questions:

$\bullet$ \textbf{RQ1: How does \ours perform as compared with state-of-the-art CF methods?}

$\bullet$ \textbf{RQ2: Does \ours exhibit any other positive effects and how does it facilitate work?}

$\bullet$ \textbf{RQ3: How do different hyper-parameter settings affect \nours?}

\begin{table}[!h]
\centering
	\caption{Statistics of the datasets.}
	\label{tab:exp-dataset}
	\begin{tabular}{c *{4}{r}}
		\toprule
		\textbf{Datasets} & \textbf{\#Users} & \textbf{\#Items} & \textbf{\#Interactions} & \textbf{Density}\\
		\midrule
            Kindle & 60,468 & 57,212 & 880,859 & 0.00025 \\
		Yelp 	& 45,477 & 30,708 & 1,777,765 & 0.00127 \\
		Books 	& 58,144 & 58,051 & 2,517,437 & 0.00075 \\
            QB-video & 30,323 & 25,730 & 1,581,136 & 0.00203 \\
		\bottomrule
	\end{tabular}
\end{table}

\begin{table*}[!ht]
    \begin{adjustbox}{max width=2.1\columnwidth}
        \begin{threeparttable}
            \centering
            \caption{Performance comparison of different recommendation models.}
            \label{tab:exp-main}
                            \begin{tabular}{ccccccccccclr}
                \toprule
                Dataset                                           & Metric    &\textbf{BPRMF} & \textbf{NeuMF}     & \textbf{NGCF} & \textbf{LightGCN}  & \textbf{SGL}              & \textbf{NCL}           & \textbf{LightGCL}   &\textbf{SimGCL}        & \textbf{DCCF}  & {~\textbf{\nours}}            & Improv.        \\ \midrule \midrule
                \multirow{6}{*}{\rotatebox{90}{Kindle}}           & Recall@10 & 0.1296        & 0.1185             & 0.1306        & 0.1570             & \underline{0.1693}        & 0.1624                 &  0.1610             & 0.1673                &  0.1450             & \textbf{0.1793}$^*$ & +5.91 \%  \\
                                                                  & Recall@20 & 0.1709        & 0.1583             & 0.1770        & 0.2080             & \underline{0.2192}        & 0.2119                 &  0.2109             & 0.2162                &  0.1956             & \textbf{0.2317}$^*$ & +5.70 \%  \\
                                                                  & Recall@50 & 0.2399        & 0.2232             & 0.2340        & 0.2888             & \underline{0.2931}        & 0.2870                 &  0.2887             & 0.2908                &  0.2770             & \textbf{0.3081}$^*$ & +5.12 \%  \\
                \cmidrule(lr){2-2}  \cmidrule(lr){3-11} \cmidrule(lr){12-13}                                                                                                                                 
                                                                  & NDCG@10   & 0.0815        & 0.0744             & 0.0796        & 0.0981             & \underline{0.1077}        & 0.1026                 &  0.1009             & 0.1067                &  0.0889             & \textbf{0.1146}$^*$ & +6.41 \%  \\
                                                                  & NDCG@20   & 0.0926        & 0.0850             & 0.0919        & 0.1117             & \underline{0.1210}        & 0.1158                 &  0.1142             & 0.1197                &  0.1023             & \textbf{0.1286}$^*$ & +6.28 \%  \\
                                                                  & NDCG@50   & 0.1073        & 0.0989             & 0.1086        & 0.1290             & \underline{0.1370}        & 0.1320                 &  0.1310             & 0.1358                &  0.1197             & \textbf{0.1451}$^*$ & +5.91 \%  \\ \midrule
                                                                                            
                \multirow{6}{*}{\rotatebox{90}{Yelp}}             & Recall@10 & 0.0643        & 0.0531             & 0.0630        & 0.0730             & 0.0833                    & 0.0920                 &  0.0743             & \underline{0.0924}    &  0.0821             & \textbf{0.0995}$^*$ & +7.68 \%  \\
                                                                  & Recall@20 & 0.1043        & 0.0885             & 0.1026        & 0.1163             & 0.1288                    & 0.1377                 &  0.1160             & \underline{0.1385}    &  0.1279             & \textbf{0.1458}$^*$ & +5.27 \%  \\
                                                                  & Recall@50 & 0.1862        & 0.1654             & 0.1864        & 0.2016             & 0.2140                    & 0.2247                 &  0.1995             & \underline{0.2264}    &  0.2179             & \textbf{0.2338}$^*$ & +3.27 \%  \\
                \cmidrule(lr){2-2}  \cmidrule(lr){3-11} \cmidrule(lr){12-13}                                                                                                             
                                                                  & NDCG@10   & 0.0458        & 0.0377             & 0.0446        & 0.0520             & 0.0601                    & 0.0678                 &  0.0543             & \underline{0.0695}    &  0.0595             & \textbf{0.0745}$^*$ & +7.19 \%  \\
                                                                  & NDCG@20   & 0.0580        & 0.0486             & 0.0567        & 0.0652             & 0.0739                    & 0.0817                 &  0.0670             & \underline{0.0834}    &  0.0736             & \textbf{0.0888}$^*$ & +6.47 \% \\
                                                                  & NDCG@50   & 0.0793        & 0.0685             & 0.0784        & 0.0875             & 0.0964                    & 0.1046                 &  0.0887             & \underline{0.1063}    &  0.0970             & \textbf{0.1121}$^*$ & +5.46 \%  \\ \midrule
                                                            
                \multirow{6}{*}{\rotatebox{90}{Books}}            & Recall@10 & 0.0607        & 0.0507             & 0.0617        & 0.0797             & 0.0898                    & 0.0933                 &  0.0842             & \underline{0.0995}    &  0.0911             & \textbf{0.1067}$^*$ & +7.24 \%  \\
                                                                  & Recall@20 & 0.0956        & 0.0823             & 0.0978        & 0.1206             & 0.1331                    & 0.1381                 &  0.1292             & \underline{0.1473}    &  0.1332             & \textbf{0.1552}$^*$ & +5.36 \%  \\
                                                                  & Recall@50 & 0.1681        & 0.0447             & 0.1699        & 0.2012             & 0.2157                    & 0.2175                 &  0.2133             & \underline{0.2315}    &  0.2095             & \textbf{0.2419}$^*$ & +4.49 \%  \\
                \cmidrule(lr){2-2}  \cmidrule(lr){3-11} \cmidrule(lr){12-13}                                                                  
                                                                  & NDCG@10   & 0.0430        & 0.0351             & 0.0427        & 0.0565             & 0.0645                    & 0.0679                 &  0.0595             & \underline{0.0725}    &  0.0663             & \textbf{0.0767}$^*$ & +5.79 \%  \\
                                                                  & NDCG@20   & 0.0537        & 0.0447             & 0.0537        & 0.0689             & 0.0777                    & 0.0815                 &  0.0731             & \underline{0.0870}    &  0.0790             & \textbf{0.0915}$^*$ & +5.17 \%  \\
                                                                  & NDCG@50   & 0.0726        & 0.0610             & 0.0725        & 0.0899             & 0.0992                    & 0.1024                 &  0.0950             & \underline{0.1092}    &  0.0990             & \textbf{0.1144}$^*$ & +4.76 \%  \\ \midrule                                                                                       
                                                                                                                        
                \multirow{6}{*}{\rotatebox{90}{QB-video}}         & Recall@10 & 0.1126        & 0.1031             & 0.1250        & 0.1264             & 0.1351                    & \underline{0.1431}     &  0.1224             & {0.1402}              &  0.1417             & \textbf{0.1485}$^*$ & +3.77 \%  \\
                                                                  & Recall@20 & 0.1762        & 0.1659             & 0.1948        & 0.1966             & 0.2044                    & \underline{0.2164}     &  0.1900             & {0.2141}              &  0.2145             & \textbf{0.2261}$^*$ & +4.48 \%  \\
                                                                  & Recall@50 & 0.2989        & 0.2851             & 0.3240        & 0.3235             & 0.3367                    & \underline{0.3515}     &  0.3157             & {0.3479}              &  0.3486             & \textbf{0.3638}$^*$ & +3.50 \%  \\
                \cmidrule(lr){2-2}  \cmidrule(lr){3-11} \cmidrule(lr){12-13}                                                                                                                      
                                                                  & NDCG@10   & 0.0790        & 0.0709             & 0.0886        & 0.0888             & 0.0946                    & \underline{0.1006}     &  0.0870             & {0.0994}              &  0.0996             & \textbf{0.1046}$^*$ & +3.98 \%  \\
                                                                  & NDCG@20   & 0.0981        & 0.0896             & 0.1094        & 0.1098             & 0.1155                    & \underline{0.1225}     &  0.1073             & {0.1215}              &  0.1216             & \textbf{0.1281}$^*$ & +4.57 \%  \\
                                                                  & NDCG@50   & 0.1305        & 0.1209             & 0.1435        & 0.1436             & 0.1504                    & \underline{0.1584}     &  0.1404             & {0.1567}              &  0.1571             & \textbf{0.1646}$^*$ & +3.91 \%  \\ \bottomrule
            
            \end{tabular}
            \begin{tablenotes}
                \item The best result is \textbf{boldfaced} and the runner-up is \underline{underlined}. $^*$ indicates the statistical significance for $p < 0.01$ compared to the best baseline.
            \end{tablenotes}
        \end{threeparttable}
    \end{adjustbox}
\end{table*}

\subsection{Experimental Setup}
\subsubsection{\textbf{Datasets}} \label{sec:dataset}

We conduct experiments on five public and real-world datasets, namely Yelp\footnote{\url{www.yelp.com/dataset}}, Books, Kindle\cite{he2016ups} (\ie Amazon-books and Amazon-kindle-store), QB-video from Tencent recommendation datasets~\cite{Tenrec}. 
These datasets originate from diverse domains, exhibiting variances in both scale and sparsity. 
For the Books and Kindle datasets, we consider interactions with ratings greater than or equal to 3 as positive feedback to transform numerical ratings into implicit feedback.
To ensure the quality of the data, we employ the 15-core setting \cite{he2016vbpr} for Yelp and Books, which ensures a minimum of 15 interactions between users and items. For the Kindle and QB-video datasets, users and items with less than 5 interactions are filtered out. Detailed statistical information about the datasets is summarized in Table~\ref{tab:exp-dataset}. We use the user-based split approach to divide the data for experimentation. 
Specifically, we divide training, validation, and test sets into a ratio of 8:1:1, respectively.
In addition, we employ pairwise sampling, where for each interaction in the training set, a negative interaction is randomly chosen to construct the training sample $(u, i^{+}, i^{-})$.

\subsubsection{\textbf{Evaluation Metrics}}
Two widely used metrics Recall@$K$ and NDCG@$K$ are adopted to quantitatively measure the recommendation performance and the values of $K$ are set to $\{10, 20, 50\}$. Following~\cite{lin2022ncl}, we adopt the all-rank protocol which ranks all candidate items that users have not interacted during evaluation.

\subsubsection{\textbf{Implementation Details}}
The proposed \textbf{\nours}, along with all baseline models, are implemented using 
RecBole\footnote{\url{https://recbole.io}} and RecBole-GNN~\cite{zhao2022recbole2}. RecBole is a unified and comprehensive framework for recommendation systems while RecBole-GNN implements various GNN-based recommendation algorithms. 
To ensure a fair comparison, we employ the same Adam~\cite{glorot2010understanding} optimizer with a learning rate of 0.001 and conduct a detailed hyperparameter search for all baseline models. The batch size is set to 4096, and the $L_2$ regularization of parameters $\lambda_2$ is chosen as $10^{-4}$. The Xavier initialization~\cite{KingmaBAdam} is utilized as the default parameter initialization method across all methods. The embedding size is set to 64.
In \nours, early stopping with the patience of 10 epochs is utilized to prevent overfitting with the indicator of NDCG@10. The hyper-parameter tuning for $\lambda_1$ is performed within the range of $\{5e^{-6}, 1e^{-6}, 5e^{-5}, 1e^{-5}\}$ and $\tau$ within the range of $\{0.05, 0.075, 0.1, 0.125, 0.15\}$. As for the models that are already implemented, we reuse the reported results from the previous work~\cite{lin2022ncl}.

\subsubsection{\textbf{Compared Methods}} \label{sec:compared-methods}
To demonstrate the effectiveness, we compare the proposed \ours with the state-of-the-art baselines with different learning paradigms: 
i) \textbf{MF}-based (BPRMF~\cite{rendle2009bpr}, NeuMF~\cite{he2017neural}). 
ii) \textbf{GNN}-based (NGCF~\cite{wang2019neural}, LightGCN~\cite{he2020lightgcn}). 
iii) \textbf{SSL}-based (SGL~\cite{wu2021self}, NCL~\cite{lin2022ncl}, LightGCL~\cite{cai2023lightgcl}, SimGCL~\cite{yu2022graph}, DCCF~\cite{ren2023disentangled}).

\begin{table*}[!ht]
    \centering
    \footnotesize
    \caption{The comparison of time complexity.}
    \label{time-complexity}
    \begin{tabular}{cccccccc}
        \toprule
        \textbf{Computation}       & \textbf{LightGCN}     & \textbf{SGL-ED}               & \textbf{NCL}                  & \textbf{LightGCL}          & \textbf{SimGCL}               & \textbf{DCCF}          & \textbf{\nours} \\
        \midrule                                                                                                                                       
        Normalization     & $O(2|E|)$    & $O((2+4\rho)|E|)$    & $O(2|E|)$            & $O(2|E|)$         & $O(2|E|)$            & $O(2|E|)$        & $O(2|E|)$\\
        SVD/Cluster       & --           & --                   & $O(2|E|Kd)$          & $O(2q|E|)$        & --                   & --               &  --   \\
        \midrule                                                                                                                                 
        Augmentation      & --           & $O(4\rho E)$         & --                   & --                & --                   & $O(2|E|Ld)$               &  --   \\
        Graph Convolution & $O(2|E|Ld)$  & $O((2+4\rho)|E|Ld)$  & $O(2|E|Ld)$          & $O(2(|E|+2qN)Ld)$ & $O(6|E|Ld)$          & $O(2(|E|+Nk)Ld )$      &  \colorbox{lightgray}{$O(2|E|d)$}  \\
        BPR Loss          & $O(2Bd)$     & $O(2Bd)$             & $O(2Bd)$             & $O(2Bd)$          & $O(2Bd)$             & $O(2Bd)$         &  $O(2Bd)$  \\
        CL Loss           & --           & $O(2(Bd + BMd))$     & $O(4(Bd + BMd))$     & $O(2L(Bd + BMd))$ & $O(2(Bd + BMd))$     & $O(6L(Bd + BMd))$ &  \colorbox{lightgray}{$O(2(Bd + BMd))$}   \\
        \bottomrule
    \end{tabular}
\end{table*}

\subsection{Overall Performance (RQ1)}
Table~\ref{tab:exp-main} lists the overall performance of all compared methods on five datasets. From 
evaluation results, we summarize the following observations:

$\bullet$ Our proposed \ours consistently outperforms all baselines across different datasets. On the Yelp and Books datasets, \ours exhibits improvement over the strongest baseline SimGCL, with an average increase of 7.46\% in Recall@10 and 6.19\% in NDCG@10. Compared to the robust baseline LightGCN, \ours shows an average improvement of 39.79\% in Recall@10 and 27.68\% in NDCG@10. These results suggest that modeling interactive collaborative signals through layer contrast in 1-hop structural neighbors can yield more informative representations and lead to enhanced performance. Furthermore, we observe that \ours performs better on sparse datasets, making it well-suited for real-world data scenarios. In dense datasets where interaction information is more abundant, other models may also optimize effectively.

$\bullet$ In the comparison of baseline models, NeuMF replaces the dot product in BPRMF with a neural network but leads to a performance decline. This may be attributed to our scenario where node representations are optimized solely based on interaction information. Consequently, more complex modeling could result in suboptimal representations. Furthermore, GNN-based models, in contrast to MF-based collaborative filtering methods, demonstrate superior performance, emphasizing the importance of utilizing graph structures to capture high-order user-item interaction signals for modeling user preferences.

$\bullet$ Across five datasets, all GCL methods significantly improve the performance of LightGCN (these GCL methods utilize multiple layers of LightGCN as the backbone). This underscores the effectiveness of integrating contrastive learning methods with collaborative filtering. Among these methods, SimGCL stands out. It maximizes the mutual information between two noise views while utilizing contrastive loss as part of the optimization objective, resulting in excellent performance. On the other hand, our \ours method simplifies the construction of contrasting pairs, innovatively introducing that we only need to perform simple layer-to-layer contrast on heterogeneous 0-th and 1-st layers, which can achieve superior performance compared to those state-of-the-art methods. 

\subsection{Complexity and Efficiency Study (RQ2)}
GCL-based collaborative filtering methods often suffer from high computational costs due to the need for constructing additional contrasting views and performing multiple graph convolutions. However, our proposed approach requires only a single execution of graph convolution and self-generated contrasting views. This significantly simplifies the computation in the training phase, making it more efficient. In this section, we give a systematic study of complexity and efficiency.  

Let $|E|$ denote the cardinality of the set of edges in the graph, $d$ represent the embedding dimension, $B$ indicate the batch size, $L$ denote the layer number of graph convolution, $M$ denote the number of nodes in a batch, $N$ represents the total number of users and items, respectively.
$\rho$ is the edge keep rate in SGL-ED (-ED is short for edge dropout)~\cite{wu2021self}.
$K$ is the number of prototype in NCL~\cite{lin2022ncl} while $q$ is the required rank in LightGCL~\cite{cai2023lightgcl}.

We compare the complexity of \ours with other competitive CL baselines. Given that none of the CL baselines introduce additional trainable parameters, they exhibit consistency in space complexity. Consequently, our emphasis shifts towards evaluating time complexity during the model training phase, as summarized in Table~\ref{time-complexity}.

$\bullet$ Some methods require pre-processing operations, for instance, SGL necessitates random augmentation on the graph, LightGCL involves singular value decomposition, and NCL requires clustering on node representations, which brings additional time costs.

$\bullet$ All GCN-based contrastive learning methods, to achieve optimal performance, typically entail multiple graph convolution computations, resulting in significant time overhead. In contrast, our method conducts contrastive learning solely on one-hop neighbors, requiring just a one-layer graph convolution, which significantly enhances model efficiency.

$\bullet$ Concerning contrastive loss computation, NCL employs two contrastive loss objectives, while LightGCL needs to calculate contrastive loss at each layer, both bringing burdens to the model's computations. In contrast, our method involves only one contrastive learning loss computation, 

In summary, \ours excels in terms of time complexity, greatly simplifying the complexity of graph-based CF models and enhancing the model's applicability to large graphs and industrial scenarios.

\begin{table}[!ht]
    \small
    \centering
    \caption{Efficiency comparison of different methods on QB-video, including average training time per epoch, number of epochs to
converge, total training time (s: second, m: minute,
h: hour).}
    \label{tab:efficiency}
    \begin{tabular}{lccccc}
      \toprule
      Method & Time/Epoch$\downarrow$ & \#Epoch$\downarrow$  & Total Time$\downarrow$ & NDCG@10$\uparrow$ \\
      \midrule
      LightGCN    & 8.8s  & 297 & 44m     &  0.0888     \\
      SGL         & 25.2s & 53  & 22m     &  0.0946     \\
      NCL         & 13.9s & 81  & 19m     &  0.1006     \\
      LightGCL    & 12.2s & 55  & 11m     &  0.0870     \\
      SimGCL      & 26.1s & 199 & 1h27m   &  0.0994     \\
      DCCF     & 101.0s & 29  & 1h11m     &  0.0996     \\
      \midrule
      \nours      & \colorbox{c1}{4.8s}  & \colorbox{c1}{49}   & \colorbox{c1}{4m}     &  \colorbox{c1}{0.1046}     \\  
      \bottomrule
    \end{tabular}
  \end{table}

In addition, we compare the training efficiency of \ours with the aforementioned models. In Table~\ref{tab:efficiency}, we present the average training time per epoch, the number of epochs to converge, and the total training time on the dataset QB-video. 
Indeed, consistent with the time complexity analysis, \ours outperforms others in a single training iteration and achieves the fastest convergence speed in practice (achieving more than 22x speedup over strong baseline SimGCL in training efficiency and convergence), which excels in superior performance while embodying simplicity and efficiency.

\subsection{Integration with Other GNN Encoders (RQ2)}
In the main experiments (Table ~\ref{tab:exp-main}), we optimize \ours with a simple one-layer LightGCN encoder. This raises the question of whether our contrastive learning object is also beneficial for other GNN-based CF encoders.
Therefore, we integrate \ours with various GNN backbones (GCN~\cite{kipfsemi}, GAT~\cite{velivckovic2017graph}, NGCF~\cite{wang2019neural}) to validate its compatibility. We obtain experimental results in Table~\ref{tab:exp-encoders}.

We can see that \ours consistently brings remarkable performance improvements compared with vanilla encoders, 
validating the superiority and adaptability of our proposed contrastive objectives. Additionally, \ours with NGCF and LightGCN as a backbone demonstrate better performance, possibly because NGCF is tailored for recommendation system scenarios. At the same time, LightGCN eliminates complex feature transformations and activation functions, ensuring more consistent representation spaces across different layers and thereby ensuring the stability of structural contrastive learning.

    \begin{table}[!h]
        \centering
        \small
        \caption{Performance comparison of different encoders when applied \nours.}
        \label{tab:exp-encoders}
        \begin{tabular}{ccccc}
          \toprule
          \multirow{2}{*}{Method} & \multicolumn{2}{c}{Yelp} & \multicolumn{2}{c}{Books} \\
          \cmidrule(lr){2-4}\cmidrule(lr){4-5}
          & Recall@10 & NDCG@10 & Recall@10 & NDCG@10 \\
          \midrule
          GCN       & 0.0573     & 0.0410     & 0.0490     & 0.0350     \\
          +\nours     & \textbf{0.0827} & \textbf{0.0599} & \textbf{0.0829} & \textbf{0.0584} \\
          \midrule
          GAT       & 0.0549     & 0.0389     & 0.0503     & 0.0352     \\
          +\nours     & \textbf{0.0890} & \textbf{0.0686} & \textbf{0.0959} & \textbf{0.0695} \\
          \midrule
          NGCF      & 0.0630     & 0.0446     & 0.0617     & 0.0427     \\
          +\nours     & \textbf{0.0917} & \textbf{0.0716} & \textbf{0.0987} & \textbf{0.0709} \\
          \midrule
          LightGCN  & 0.0730     & 0.0520     & 0.0797     & 0.0565     \\
          +\nours     & \textbf{0.0995} & \textbf{0.0745} & \textbf{0.1067} & \textbf{0.0767} \\
          \midrule
          Avg Improv. & 47.07\% & 56.56\% & 63.42\% & 66.52\% \\
          \bottomrule
        \end{tabular}
      \end{table}

\subsection{Resistance Against Data Sparsity (RQ2)}
To further validate the resistance of \ours in mitigating data sparsity, we divide all users into five groups based on their interaction frequencies while keeping the same total interaction count in each group. We then calculate Recall@10 for LightGCN and \ours on these five user groups. 

\begin{figure}[!ht]
	{
		\begin{minipage}[t]{0.49\linewidth}
			\centering
			\includegraphics[width=1\textwidth]{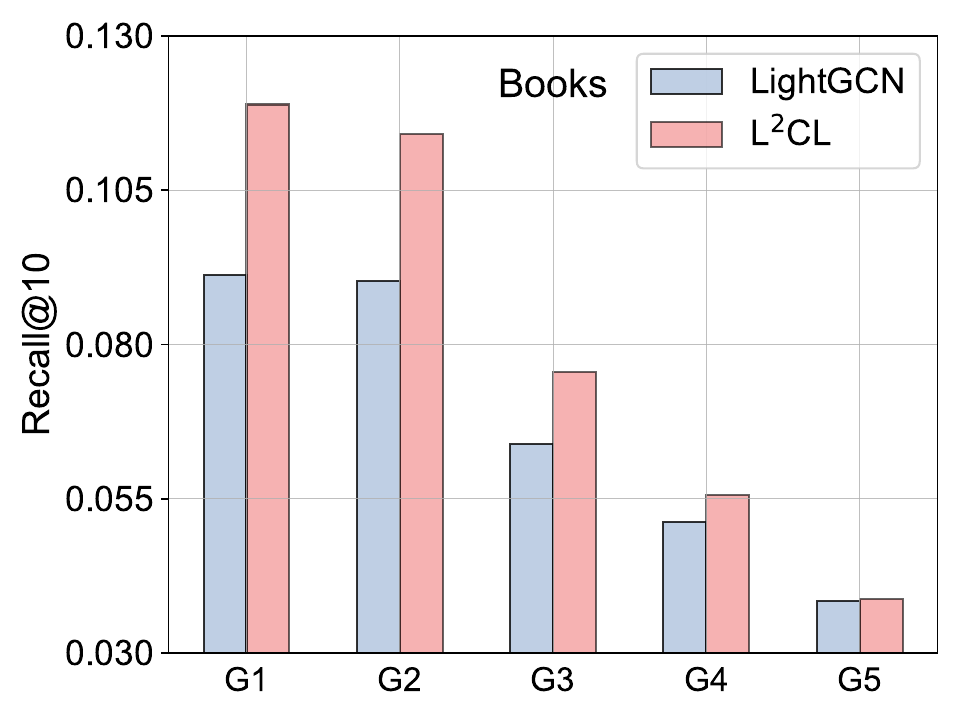}
		\end{minipage}
		\begin{minipage}[t]{0.49\linewidth}
			\centering
			\includegraphics[width=1\textwidth]{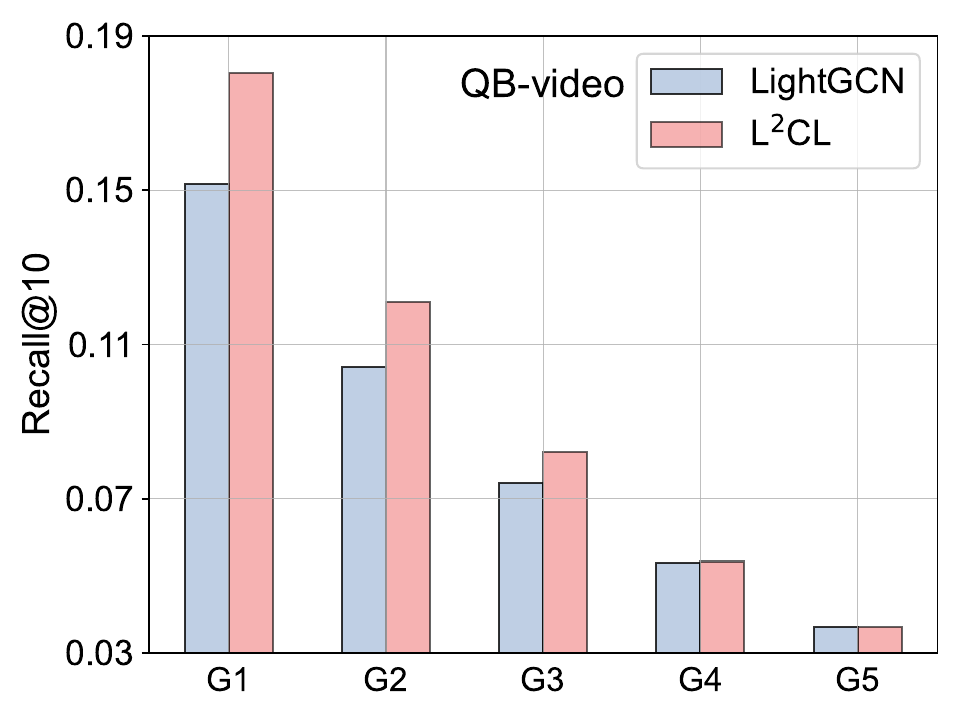}
		\end{minipage}
	}
	\caption{Performance on users of different sparsity degrees in terms of Recall@10. G1 denotes the group of users with the lowest average number of interactions.
 }
	\label{fig: sparsity}

\end{figure}

The results, as shown in Figure~\ref{fig: sparsity}, indicate that the performance of \ours consistently surpasses LightGCN. Additionally, as the interaction frequency decreases, the performance gain from \ours increases, which implies the robustness of \ours to effectively alleviate the sparsity of interaction data. \ours performs best on the most sparse user group (G1), which can be attributed to our explicit utilization of 1-hop neighbors in contrastive learning, bringing more powerful structure-awareness ability. 

\subsection{Visualization Analysis (RQ2)}
\begin{figure}[!ht]
	{
		\begin{minipage}[t]{0.495\linewidth}
			\centering
			\includegraphics[width=1\textwidth]{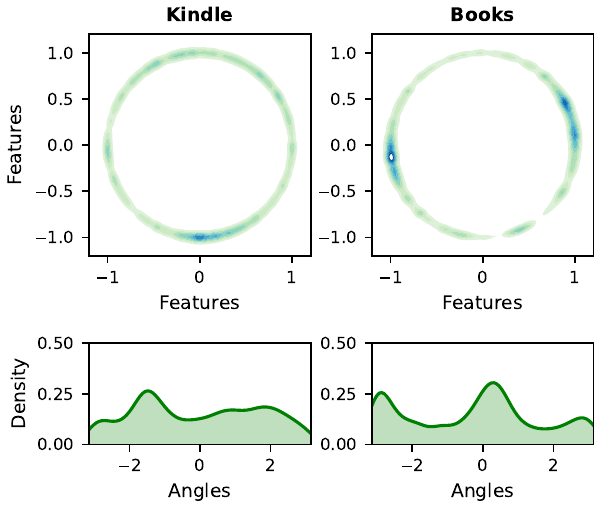}
                \subcaption{LightGCN}
		\end{minipage}
		\begin{minipage}[t]{0.495\linewidth}
			\centering
			\includegraphics[width=1\textwidth]{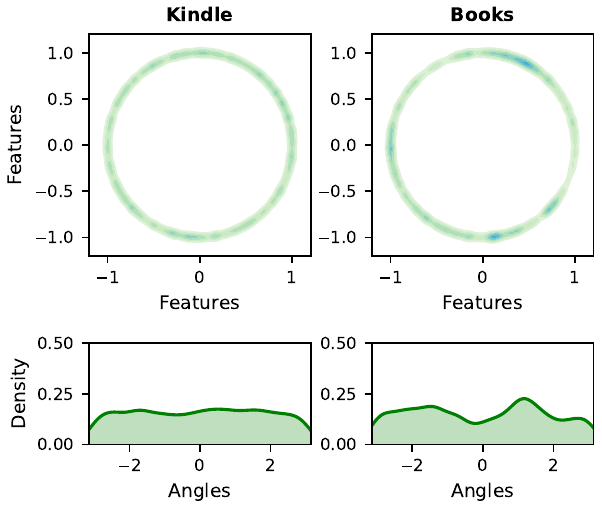}
                \subcaption{\nours}
		\end{minipage}
	}
	\caption{User distributions with Gaussian kernel density estimation (KDE) in $\mathcal{R}^{2}$ (the darker the color is, the more points fall in that area.) and KDE on angles (\ie arctan2(y, x) for each point (x,y) $\in S^1$).}
	\label{fig:kde-whole}
\end{figure}

In this section, we demonstrate the effectiveness of \ours in learning a moderately dispersed representation distribution by introducing informative variance to maximize the benefits derived from CL. Specifically, we randomly sample 2,000 users from Kindle, Books
respectively, and map the learned embeddings (when the methods reach their best performance) to 2-dimensional normalized vectors on the unit hypersphere with t-SNE~\cite{van2008visualizing}. 
Then we plot the feature distributions and density estimations on angles for each point with Gaussian kernel density estimation (KDE)~\cite{chen2017tutorial} in Figure~\ref{fig:kde-whole}.

From Figure~\ref{fig:kde-whole}, we can see that the representations learned by LightGCN are clustered on several clusters, while the representations learned by \ours are more uniform. As mentioned in previous works~\cite{wang2020understanding, yu2022graph, wang2022towards}, a more uniform distribution allows the learned representation to have better intrinsic characteristics while preferring to keep as much information as possible, leading to a better ability for preserving unique user preference.

\subsection{Hyperparameter Sensitivity Analysis (RQ3)}
This section investigates the sensitivity of several key hyperparameters on the recommendation performance of \nours. The evaluation results in terms of Recall@10 are presented in
Figure~\ref{fig:hyper-tau}, ~\ref{fig:hyper-alpha}, ~\ref{fig:hyper-lambda}. 
\subsubsection{\textbf{Performance Comparison w.r.t. \boldsymbol{$\tau$}}}
\begin{figure}[!h]
	{
		\begin{minipage}[t]{0.325\linewidth}
			\centering
			\includegraphics[width=1\textwidth]{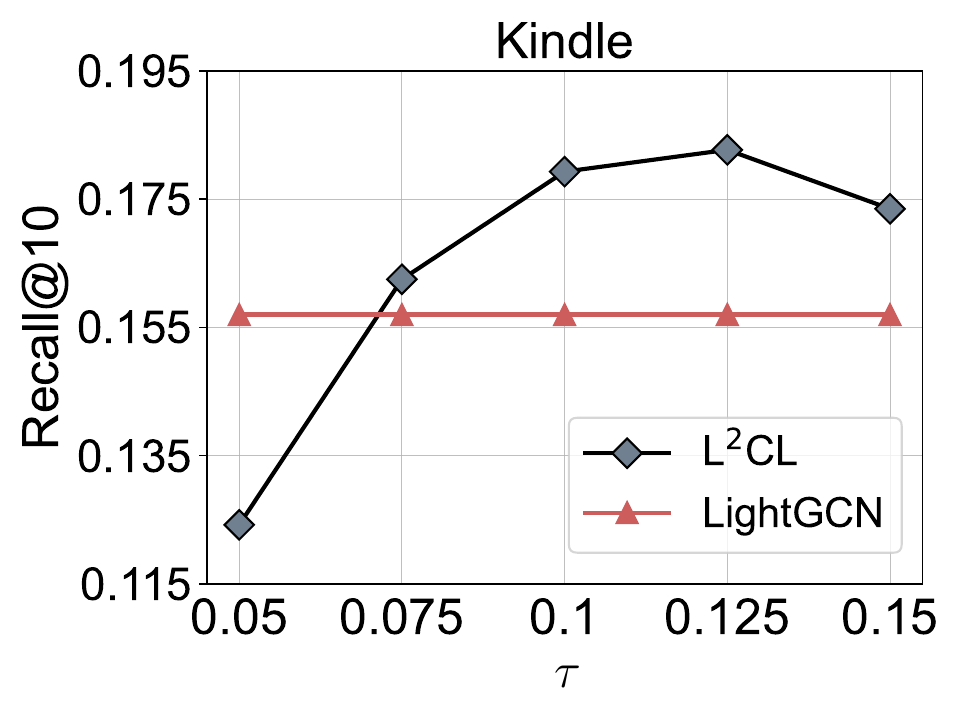}
		\end{minipage}
		\begin{minipage}[t]{0.325\linewidth}
			\centering
			\includegraphics[width=1\textwidth]{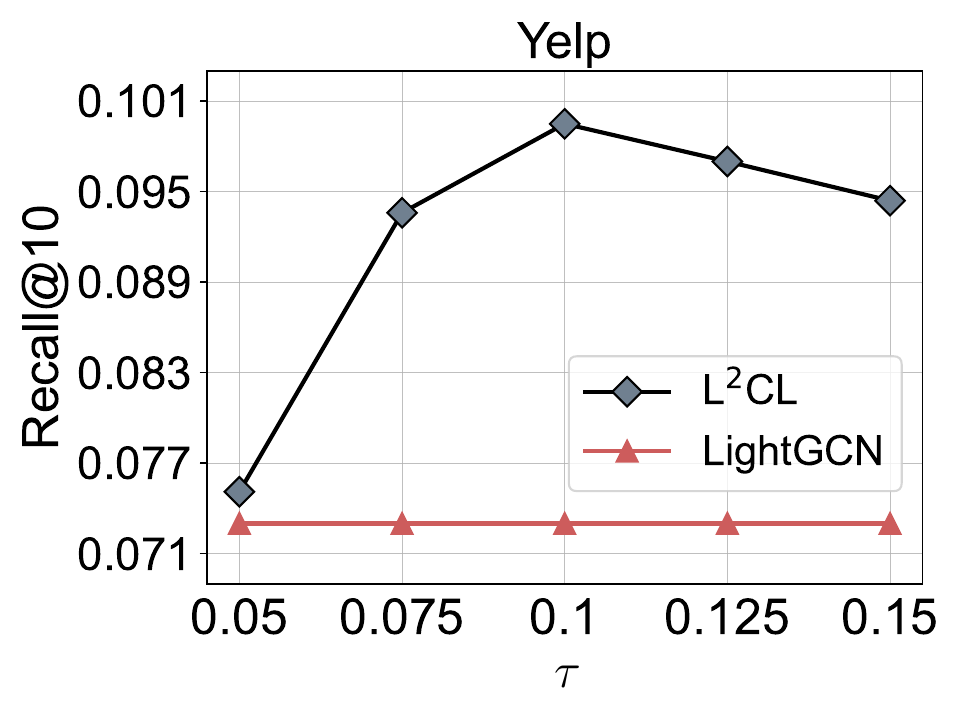}
		\end{minipage}
		\begin{minipage}[t]{0.325\linewidth}
			\centering
			\includegraphics[width=1\textwidth]{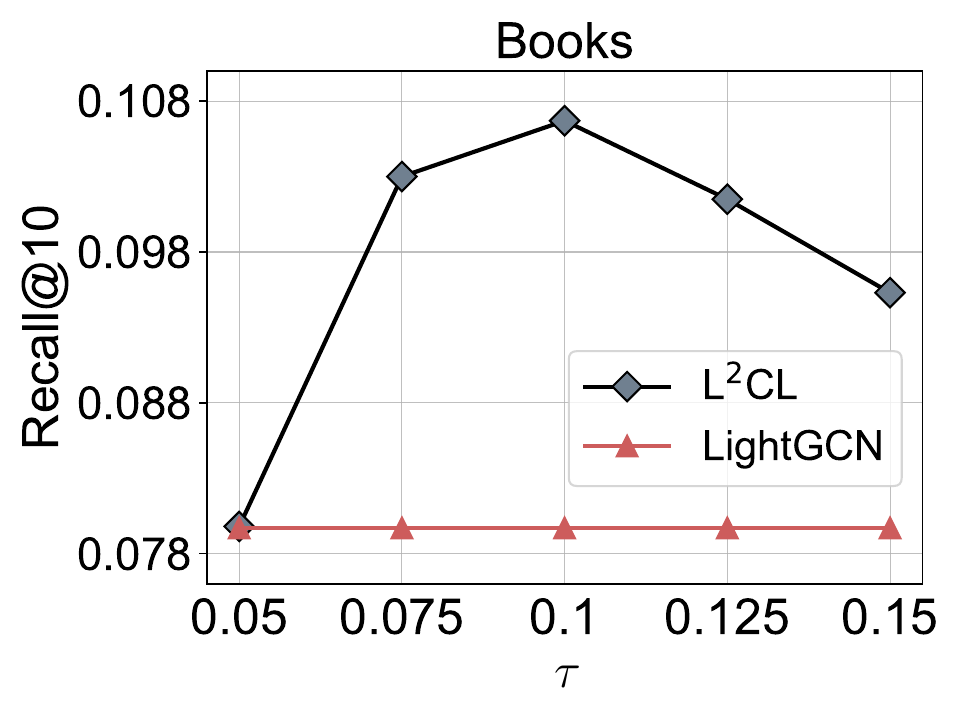}
		\end{minipage}
	}
	\caption{Performance analysis (Recall@10) for different $\tau$ on LightGCN and \nours.}\label{fig:hyper-tau}
\end{figure}

$\tau$ denotes the tunable temperature hyperparameter to adjust the scale for softmax in InfoNCE~\cite{oord2018representation}, which determines the level of attention the contrastive loss pays to hard negative samples~\cite{wang2021understanding}. From Figure~\ref{fig:hyper-tau}, 
both excessively small and large values of $\tau$ will decrease the model's performance, which is associated with previous study~\cite{wang2021understanding}. Specifically, a too-lower temperature setting (\eg $\tau=0.05$) will over-emphasize the gradient contributions of hard negative samples which are usually false negatives, making it challenging to align the anchor with its positive neighborhood nodes used in our 1-hop contrastive learning in embedding space.
In practice, setting $\tau=0.1$ tends to lead to optimal model performance under typical circumstances.

\subsubsection{\textbf{Performance Comparison w.r.t. \boldsymbol{$\alpha$}}}
\begin{figure}[!h]
	{
		\begin{minipage}[t]{0.325\linewidth}
			\centering
			\includegraphics[width=1\textwidth]{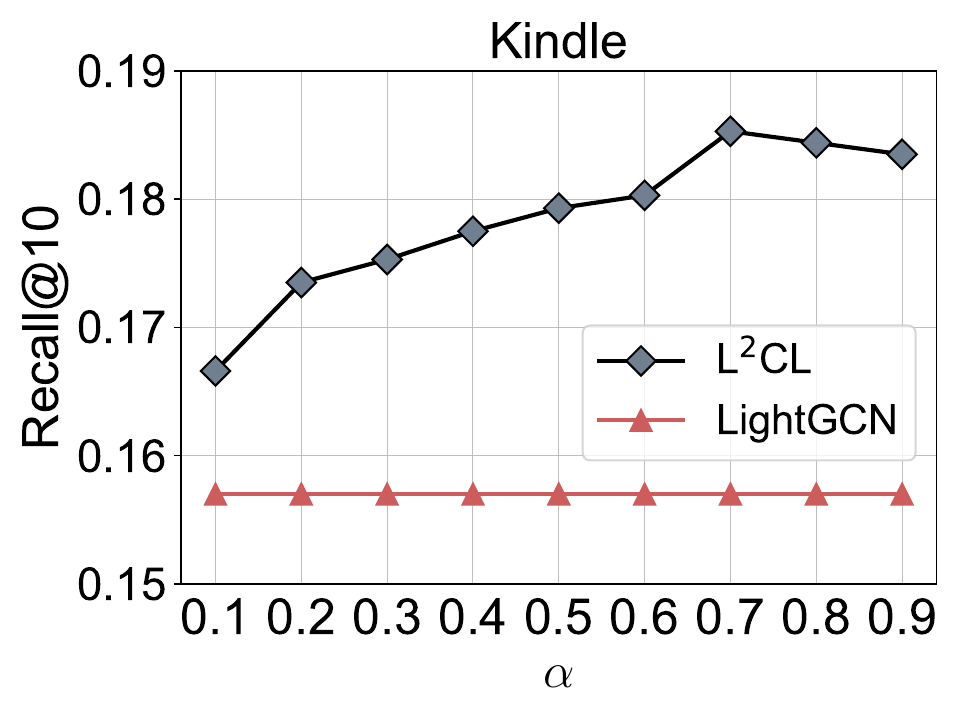}
		\end{minipage}
		\begin{minipage}[t]{0.325\linewidth}
			\centering
			\includegraphics[width=1\textwidth]{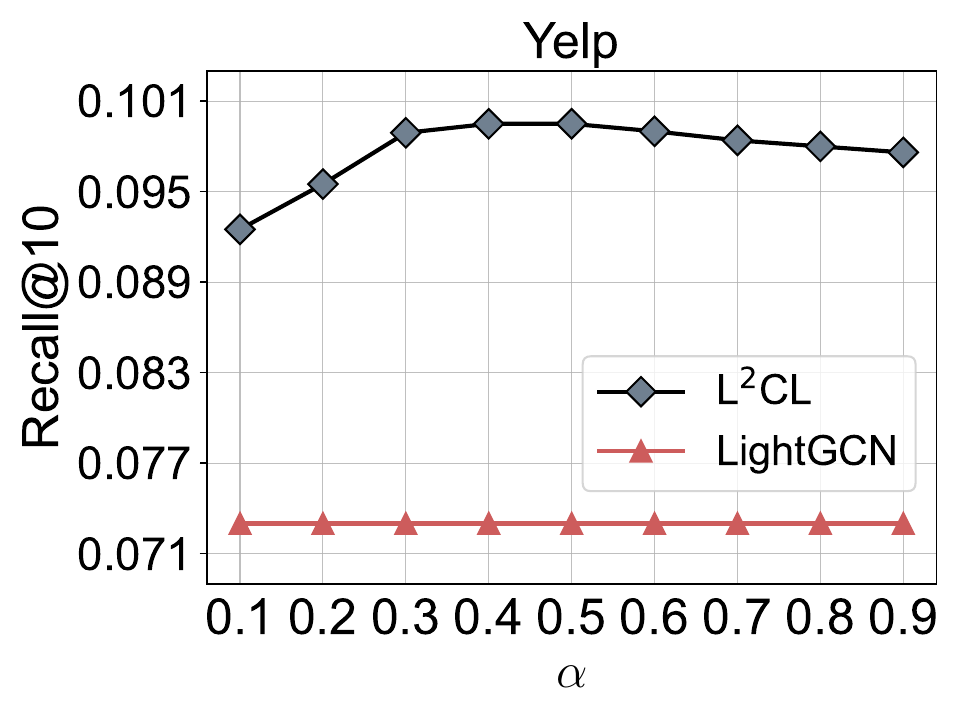}
		\end{minipage}
		\begin{minipage}[t]{0.325\linewidth}
			\centering
			\includegraphics[width=1\textwidth]{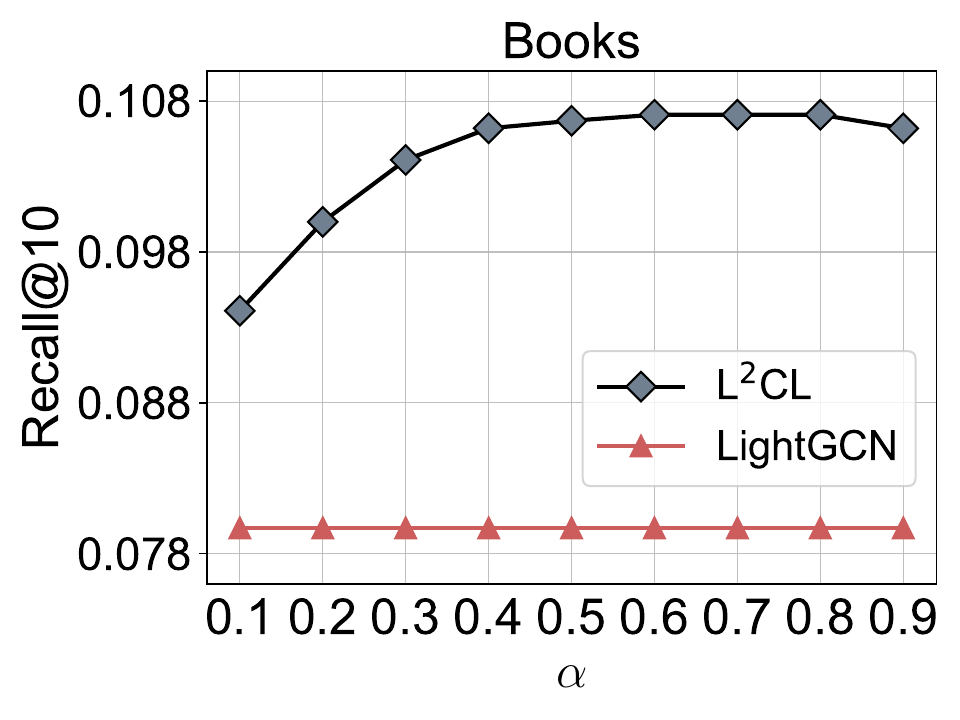}
		\end{minipage}
	}
	\caption{Performance analysis (Recall@10) for different $\alpha$ on LightGCN and \nours.}\label{fig:hyper-alpha}
\end{figure}

The coefficient $\alpha$ balances the two types of contrastive learning losses in Eq. (\ref{cl_loss_whole}). 
As shown in Figure~\ref{fig:hyper-alpha}, \ours is consistently outperforming
LightGCN, demonstrating the robustness of our method to parameter $\alpha$.
Specifically, for the Kindle dataset, the model performs better when user-side representation learning is emphasized. This observation might be attributed to that Kindle is more sparse, where the limited interactions may not be sufficient to fully capture user preferences for items. 
Therefore, the model advocates assigning greater weight to user-side representation learning to capture this missing information.
Overall, utilizing a value of $0.5$ to $\alpha$ is able to achieve competitive recommendation performance, which indicates the importance of learning both enhanced user and item representation. 
\subsubsection{\textbf{Performance Comparison w.r.t. $\boldsymbol{\lambda_1}$}}
\begin{figure}[!h]
	{
		\begin{minipage}[t]{0.325\linewidth}
			\centering
			\includegraphics[width=1\textwidth]{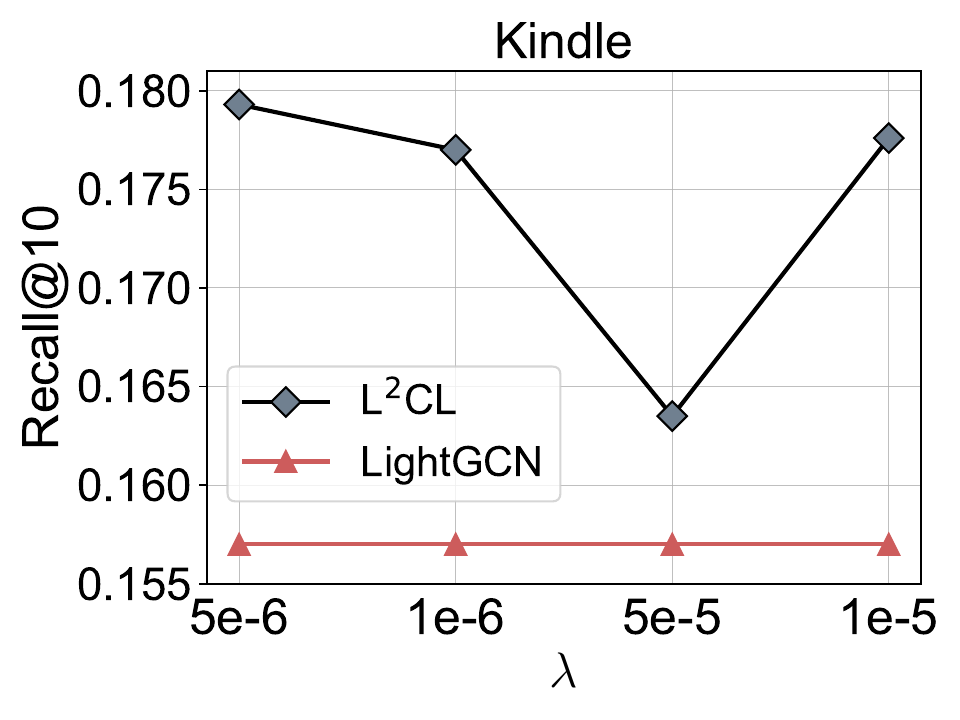}
		\end{minipage}
		\begin{minipage}[t]{0.325\linewidth}
			\centering
			\includegraphics[width=1\textwidth]{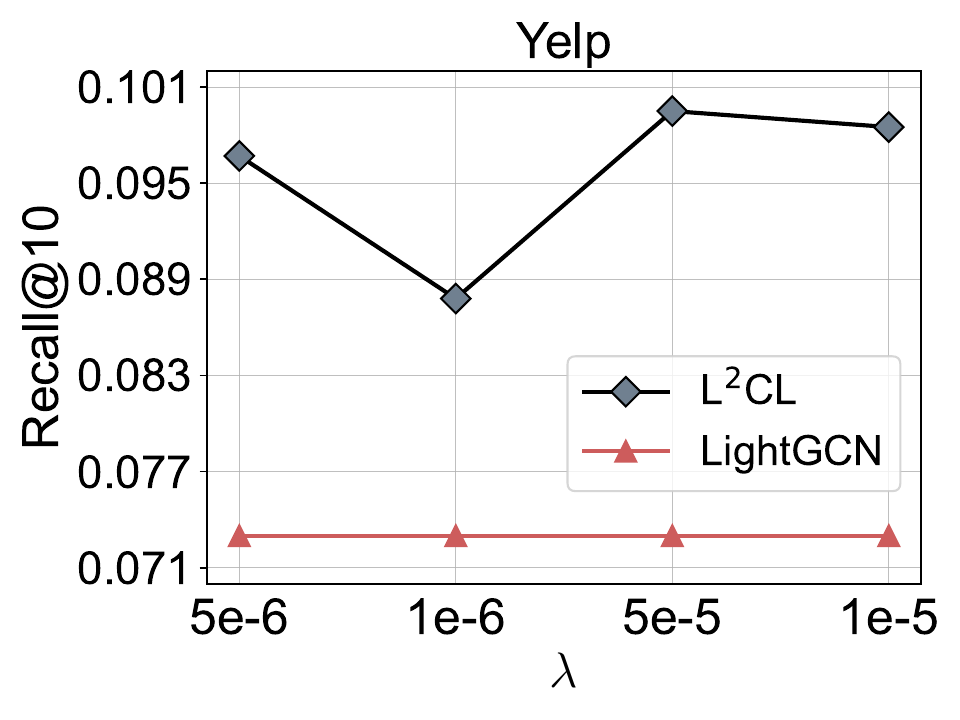}
		\end{minipage}
		\begin{minipage}[t]{0.325\linewidth}
			\centering
			\includegraphics[width=1\textwidth]{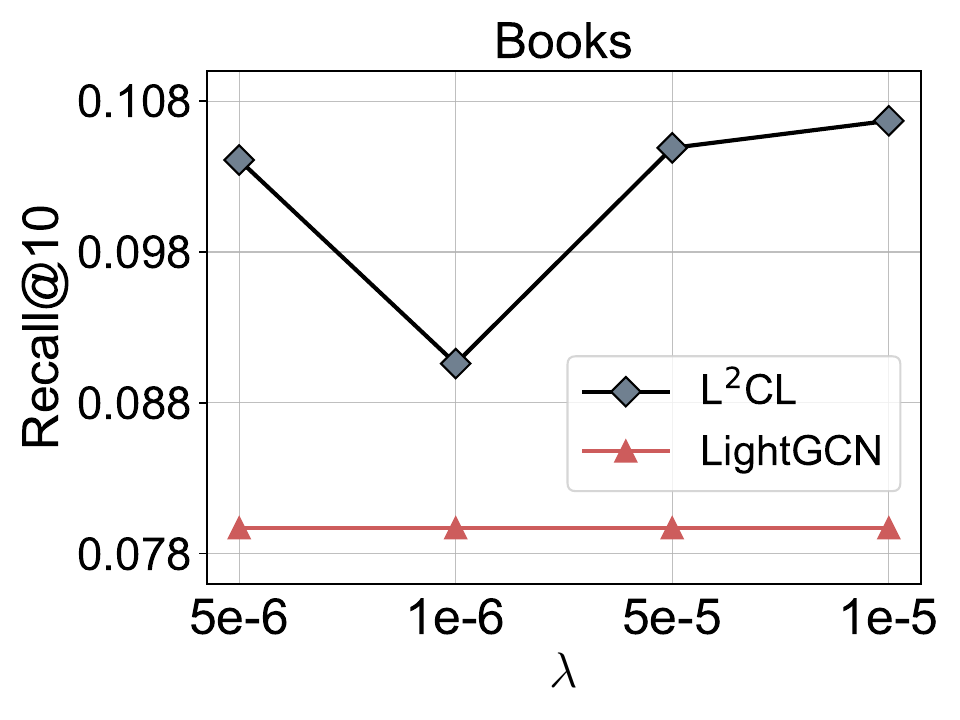}
		\end{minipage}
	}
	\caption{Performance analysis (Recall@10) for different $\lambda_1$ on LightGCN and \nours.}\label{fig:hyper-lambda}
\end{figure}

$\lambda_1$ controls the strength of contrastive learning. Specifically, the weight $\lambda_1$ is searched in the range of \{$5e^{-6}, 1e^{-6}, 5e^{-5}, 1e^{-5}$\} to explore its impact on the model’s performance. The results are presented in Figure ~\ref{fig:hyper-lambda}. It is observed that the best performance is achieved nearly $1e^{-5}$, which suggests that a reasonable $\lambda_1$ 
could balance contrastive learning loss and recommendation loss, leading to better optimization.
\section{Related Work}
\subsection{Graph-based Collaborative Filtering}
Collaborative filtering (CF) is a widely used algorithm in recommendation systems.
The key is to learn the representation of users/items through capturing collaborative signals on historical interactions~\cite{sarwar2001item, koren2008factorization}.
Early CF algorithms leverage matrix factorization~\cite{koren2009matrix, mnih2007probabilistic} to map user's and item's IDs to latent embeddings and use inner product or neural networks~\cite{rendle2009bpr, he2017neural} to estimate the preferences.

Recently, there have been many research efforts focused on deploying graph representation learning to capture high-order connectivity between users and items and modeling complex pairwise relationships.
Early studies~\cite{gori2007itemrank, yang2018hop} utilize label propagation and random walk to capture graph structure information. Most recent works directly use graph neural networks  (GNNs)~\cite{kipfsemi, wu2019simplifying} to enhance collaborative filtering.
Pinsage~\cite{ying2018graph} and NGCF~\cite{wang2019neural} introduce graph convolution in recommender to exploit higher-order connectivity for improved performance. DGCF~\cite{wang2020disentangled} constructs disentangled representations for users and items.
Additionally, certain methods~\cite{mao2021ultragcn, he2020lightgcn, jin2023enhancing} aim to enhance the efficiency and performance of GNNs by streamlining the architecture and altering the way of embedding propagation. Notably, by eliminating redundant operations, such as the transformation matrix and activation function, LightGCN~\cite{he2020lightgcn} has emerged as the most popular GNN collaborative filtering method due to its simplicity and robust performance.

\subsection{Contrastive Learning for Recommendation}
As a prevalent self-supervised learning paradigm, contrastive learning (CL) has demonstrated impressive representation learning capabilities with limited labeled data in vision and language research~\cite{oord2018representation, tian2020makes, gao2021simcse}. 
In essence, CL generates diverse views and subsequently maximizes the consistency among two contrastive views to preserve invariant representations. 
Recent studies~\cite{yao2021, wu2021self, lin2022ncl, cai2023lightgcl, yu2022graph, yu2023xsimgcl, jiang2023adaptive} have successfully deployed contrastive learning to graph collaborative filtering scenarios to tackle challenges such as data sparsity and noise.
For instance, SGL~\cite{wu2021self} generates contrasting views of the user-item interaction graph by employing graph augmentations like edge-dropout operations. NCL~\cite{lin2022ncl} promotes representation learning by jointly contrasting neighbors from the graph structure and semantic distribution. 
LightGCL~\cite{cai2023lightgcl} employs singular value decomposition to extract local-global collaborative filtering signals. 
In addition, SimGCL~\cite{yu2022graph} and XSimGCL~\cite{yu2023xsimgcl} introduce uniform distribution noise to the representations and enhance the consistency of embedding learning across two contrastive views. 

However, current works rely on the graph augmentation strategies, which are primarily designed based on empirical intuition, heuristics, and experimental trial-and-error.
This lack of generalizability makes it challenging to remain robust against noise perturbations across different datasets and downstream tasks. 
In this work, we propose to unify structural contrastive learning to a layer-to-layer framework, which provides guidance for simplifying and efficiently improving representation quality for recommendation.
\section{Conclusion}
In this work, we revisit GCL-based recommendation methods from the perspective of layer-to-layer contrasting. Firstly, we demonstrate that current GCL-based methods generally follow a layer-to-layer contrasting paradigm, where representations learned from different layers (hops) naturally form two contrastive views without additional augmentations.
Based on our understanding, we propose \nours, an augmentation-free layer-to-layer contrastive learning framework to unify the GCL learning paradigm for recommendations.
We then conduct empirical studies on \ours and discover that contrastive representations from shallow layers can achieve better performance. We analyze possible reasons to explain these findings and provide theoretical results that support our observations. 
Our analysis further motivates us to simplify \ours with an embarrassingly simple one-hop layer-to-layer contrasting scheme. In our experiments, \ours exhibits significantly improved performance over various state-of-the-art methods and performs better in efficiency and robustness against data sparsity.

\balance
\bibliographystyle{ACM-Reference-Format}
\bibliography{ref}

\end{document}